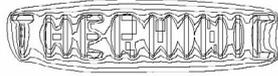

# Very Fast Chip-level Thermal Analysis

Keiji Nakabayashi[†], Tamiyo Nakabayashi[‡], and Kazuo Nakajima[*]

[†]Graduate School of Information Science, Nara Institute of Science and Technology
Keihanna Science City, Nara, Japan, keiji-n@is.naist.jp
[‡]Graduate School of Humanities and Sciences, Nara Women's University
Kitauoyahigashi-machi, Nara, Japan, nakaba@ics.nara-wu.ac.jp
[*]Dept. of Electrical and Computer Engineering, University of Maryland
College Park, MD 20742, USA, nakajima@umd.edu

**Abstract**

We present a new technique of VLSI chip-level thermal analysis. We extend a newly developed method of solving two dimensional Laplace equations to thermal analysis of four adjacent materials on a mother board. We implement our technique in C and compare its performance to that of a commercial CAD tool. Our experimental results show that our program runs 5.8 and 8.9 times faster while keeping smaller residuals by 5 and 1 order of magnitude, respectively.

## 1. Introduction

As technology advances towards the physical limits of VLSI devices, a variety of physical phenomena play much heavier roles in their performances. For example, heat generation and thermal conduction within silicon substrate and from metal layers have innegligible effects on the characteristics of VLSI chips in the post-90 nm era [5]. Such thermal phenomena are also an important factor for system design on a mother board [4] [12] [16].

Among main research activities on thermal analysis [9] [15] is the modeling of thermal conduction by a Laplace equation and its solution by such methods as finite difference (FDM), finite element (FEM), and boundary element methods (BEM) [10] [12] [16]. As such a number of CAD tools for thermal analysis are commercially available. In fact, Raphael from Synopsis was developed as a Laplace equation solver and has successfully been used in many relevant areas [11]. This software uses FDM to discretize a Laplace equation and solves the resultant huge system of linear equations by Incomplete Cholesky Conjugate Gradient method (ICCG) [11].

Indirect or iterative methods such as ICCG can produce numerical solutions with less accuracy for large scale systems of linear equations faster than direct methods such as LU decomposition. Recently, however, a new efficient direct method, called Symbolic Partial Solution Method (S-PSM) was introduced in the area of computational fluid dynamics [2] [3]. This method was derived from the general linear equation solver, called Partial Solution Method (PSM) [8], and is applicable to the case of tridiagonal coefficient matrices. S-PSM was later applied in a Laplace equation solution process for thermal conduction analysis of two adjacent materials [6].

We extend this S-PSM-based Laplace equation solver and present a new chip-level thermal analysis technique. We consider a realistic case of four layer materials of different thermal conductivities on a mother board. We first describe thermal conduction in these materials by a set of Laplace equations. We then use FDM to discretize the equations and generate large scale systems of linear equations. Finally we apply S-PSM to get numerical solutions for the linear systems. We implemented this technique in C and ran the program and Raphael on the four layer case. We show that our program generated solutions (1) at 5.8 and 8.9 times faster speeds, and (2) with 5 and 1 order of magnitude smaller residuals, respectively, than Raphael.

In the next section, we explain how S-PSM works using our four layer example. Section III describes our experiments and compares the results. We conclude the paper in Section IV.

## 2. The Problem

Consider a multi-layer structure as depicted in Fig. 2.1, where four layers of materials $p$, $q$, $r$, and $s$ of thermal conductivities $k_p$, $k_q$, $k_r$, and $k_s$, respectively, are stacked together. This structure may represent portion of a mother board design, where the layers are from the bottom, a VLSI chip, a thermal interface material (*e.g.*, adhesive), a heat spreader, and a heat sink. Heat is generated from the heat source of active silicon bulk, which is located at the bottom of the VLSI chip. The heat travels through a heat transfer pass consisting of the chip die, the adhesive, the heat spreader, and the heat sink, and goes out to the ambient air. Our problem is to find temperature distribution through two dimensional steady-state thermal conduction analysis.

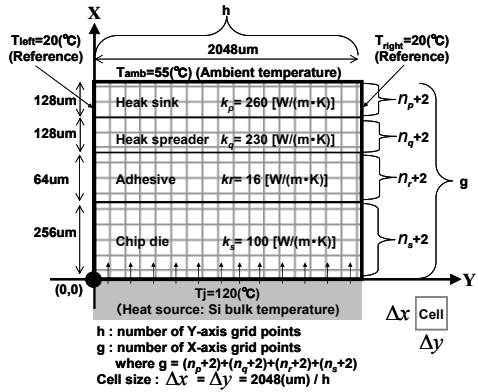

**Fig. 2.1: Two dimensional analysis for steady-state heat conduction of four materials.**

## 3. S-PSM-based Solution Process

We describe the concept and a computational procedure of the linear equation solver, called S-PSM as applied to Laplace equations. Given a system of linear equations derived by FDM, the S-PSM decomposes it into its subsystems and finds the values of the variables shared by each pair of adjacent subsystems. Figure 3.1 shows an overall flow of the major algebraic computations to take place at each subsystem with their relevant equations and solutions specified.

It should be noted from the figure that the S-PSM-based solution process goes through many levels of repeated operations of decomposition and merging. In the following section, the level information is attached to variable vectors and coefficient submatrices as their superscripts with parentheses such as (0) and (e).

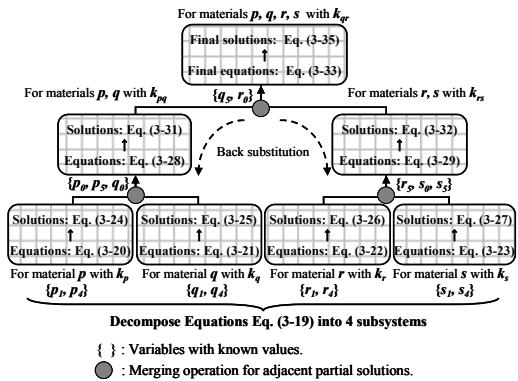

**Fig. 3.1: An overall computation flow with relevant equation numbers of a S-PSM-based solution process**.

### 3.1. Laplace and Finite Difference Equations

Heat diffusion is described by the Laplace equation

$$k_u(\frac{\partial^2 u}{\partial x^2}+\frac{\partial^2 u}{\partial y^2}) = 0 \quad (3\text{-}1)$$

where $u$ is the variable associated with the material $u$ under consideration and $k_u$ represents its thermal conductivity. In our analysis, the bold lower case letter $u$ represents one of the four materials $p$, $q$, $r$ and $s$ and the lower case letter $u$ denotes one of their corresponding variable names $p$, $q$, $r$ and $s$. So we have a set of four Laplace equations to solve.

When we apply FDM to Eq. (3-1), we decompose each of the corresponding rectangle domains into a grid of $(n_u+2) \times m$ points to which variables are assigned as shown inside each rectangle of Fig. 3.2. Furthermore, two sets of $(n_u+2)$ variables are assigned to the boundary points on the left and right borders and two sets of $m$ variables to those on the upper and lower borders of each rectangle in the figure. They all represent the boundary conditions to be described later.

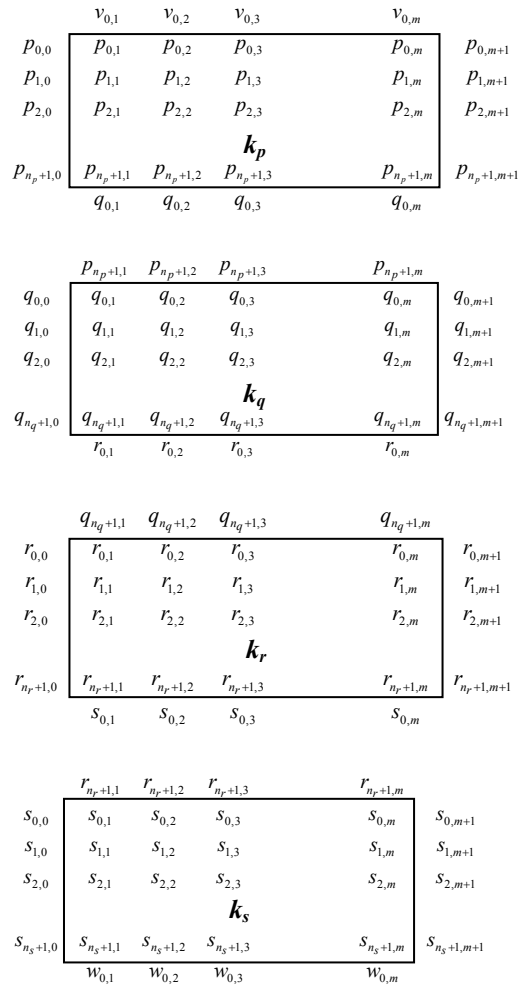

**Fig. 3.2: The arrangement of interior grid and boundary points for the four material domains.**

For the sake of simplicity, we make two assumptions:
(1) Each cell of the four grids is a square of the same size, that is, $\Delta x = \Delta y$.
(2) The number of X-axis grid points for each material $u$ (=$p$, $q$, $r$, $s$) is a power of 2, that is,
$$n_p = 2^{e_1+2}, n_q = 2^{e_2+2}, n_r = 2^{e_3+2}, n_s = 2^{e_4+2} \quad (3\text{-}2)$$

for some positive integers $e_1$, $e_2$, $e_3$ and $e_4$, respectively.

In the following, whenever two indices $i$ and $j$ are used as subscripts for interior grid point variables and their relevant vectors, their respective ranges are $i = 1, 2, \ldots, n_u$ and $j = 1, 2, \ldots, m$. As the thermal conductivity is the same at each grid point in each domain, the application of FDM yields

$$k_u\left(\frac{u_{i-1,j} - 2u_{i,j} + u_{i+1,j}}{\Delta x^2} + \frac{u_{i,j-1} - 2u_{i,j} + u_{i,j+1}}{\Delta y^2}\right) = 0 \quad (3\text{-}3)$$

As $\Delta x = \Delta y$, we derive

$$u_{i-1,j} + u_{i+1,j} + u_{i,j-1} + u_{i,j+1} - 4u_{i,j} = 0 \quad (3\text{-}4)$$

### 3.2. Boundary Conditions

As shown in Fig.2.1, the upper and lower boundary conditions are the ambient temperature $T_{amb}$ of 55℃ and the heat source temperature $T_j$ of 120 ℃. Likewise the left and right boundary conditions are the reference temperatures $T_{left}$ and $T_{righ}$ of 20℃ each.

The left and right boundary conditions for each material $u$ are expressed in the equations as

$$\mathbf{f}_{u,i}^{(0)} = (-u_{i,0} \quad 0 \quad 0 \quad \ldots \quad 0 \quad -u_{i,m+1})^T \quad (3\text{-}5)$$

The top and bottom boundary conditions for the combined four layers of materials are given as

$$\mathbf{v}^{(0)} = (v_{0,1} \quad v_{0,2} \quad \ldots \quad v_{0,m-1} \quad v_{0,m})^T \quad (3\text{-}6)$$

$$\mathbf{w}^{(0)} = (w_{0,1} \quad w_{0,2} \quad \ldots \quad w_{0,m-1} \quad w_{0,m})^T \quad (3\text{-}7)$$

Note that the superscripts (0) for the above vectors indicate that their element values are given at the start of the S-PSM process.

We set the thermal conductivity of the boundary between adjacent materials, say $q$ and $r$ of thermal conductivities $k_q$ and $k_r$ as

$$k_{q,r} = \frac{2k_q k_r}{k_q + k_r} \quad (3\text{-}8)$$

At the boundary between each pair of materials, we use the first order approximation for heat conduction as follows. From the viewpoint of material $q$, the following difference equation holds at its boundary with material $p$:

$$k_{p,q}(p_{n_p+1,j} - q_{0,j}) + k_q(q_{0,j-1} - q_{0,j}) + k_q(q_{0,j+1} - q_{0,j}) + k_q(q_{1,j} - q_{0,j}) = 0 \quad (3\text{-}9)$$

Similarly, at its boundary with material $r$, we have

$$k_q(q_{n_q,j} - q_{n_q+1,j}) + k_q(q_{n_q+1,j-1} - q_{n_q+1,j}) + k_q(q_{n_q+1,j+1} - q_{n_q+1,j}) + k_{q,r}(r_{0,j} - q_{n_q+1,j}) = 0 \quad (3\text{-}10)$$

Thus, using the variable vector notation

$$\mathbf{u}_i^{(0)} = (u_{i,1} \quad u_{i,2} \quad u_{i,3} \quad \ldots \quad u_{i,m-1} \quad u_{i,m})^T \quad (3\text{-}11)$$

we have the following four equations for the pair of adjacent materials, $q$ and $r$.

$$\left.\begin{array}{l} k_{p,q}\mathbf{p}_{n_p+1}^{(e_1)} + A_{q1}^{(0)}\mathbf{q}_0^{(e_2)} + k_q\mathbf{q}_1^{(e_2)} = k_q\mathbf{f}_{q,0}^{(0)} \\ k_q\mathbf{q}_{n_q}^{(e_2)} + A_{q2}^{(0)}\mathbf{q}_{n_q+1}^{(e_2)} + k_{q,r}\mathbf{r}_0^{(e_3)} = k_q\mathbf{f}_{q,n_q+1}^{(0)} \\ k_{q,r}\mathbf{q}_{n_q+1}^{(e_2)} + A_{r1}^{(0)}\mathbf{r}_0^{(e_3)} + k_r\mathbf{r}_1^{(e_3)} = k_r\mathbf{f}_{r,0}^{(0)} \\ k_r\mathbf{r}_{n_r}^{(e_3)} + A_{r2}^{(0)}\mathbf{r}_{n_r+1}^{(e_3)} + k_{r,s}\mathbf{s}_0^{(e_4)} = k_r\mathbf{f}_{r,n_r+1}^{(0)} \\ \mathbf{p}_{n_p+1}^{(e_1)} = \mathbf{p}_5^{(e_1)}, \quad \mathbf{q}_{n_q}^{(e_2)} = \mathbf{q}_4^{(e_2)}, \quad \mathbf{q}_{n_q+1}^{(e_2)} = \mathbf{q}_5^{(e_2)} \\ \mathbf{r}_{n_r}^{(e_3)} = \mathbf{r}_4^{(e_3)}, \quad \mathbf{r}_{n_r+1}^{(e_3)} = \mathbf{r}_5^{(e_3)} \end{array}\right\} \quad (3\text{-}12)$$

where

$$A_{q1}^{(0)} = \begin{pmatrix} -(k_{p,q}+3k_q) & k_q & & & \\ k_q & -(k_{p,q}+3k_q) & k_q & & \\ & k_q & -(k_{p,q}+3k_q) & k_q & \\ & & \bullet & \bullet & \bullet \\ & & & k_q & -(k_{p,q}+3k_q) \end{pmatrix} \quad (3\text{-}13)$$

$$A_{q2}^{(0)} = \begin{pmatrix} -(3k_q+k_{q,r}) & k_q & & & \\ k_q & -(3k_q+k_{q,r}) & k_q & & \\ & k_q & -(3k_q+k_{q,r}) & k_q & \\ & & \bullet & \bullet & \bullet \\ & & & k_q & -(3k_q+k_{q,r}) \end{pmatrix} \quad (3\text{-}14)$$

$$A_{r1}^{(0)} = \begin{pmatrix} -(k_{q,r}+3k_r) & k_r & & & \\ k_r & -(k_{q,r}+3k_r) & k_r & & \\ & k_r & -(k_{q,r}+3k_r) & k_r & \\ & & \bullet & \bullet & \bullet \\ & & & k_r & -(k_{q,r}+3k_r) \end{pmatrix} \quad (3\text{-}15)$$

$$A_{r2}^{(0)} = \begin{pmatrix} -(3k_r+k_{r,s}) & k_r & & & \\ k_r & -(3k_r+k_{r,s}) & k_r & & \\ & k_r & -(3k_r+k_{r,s}) & k_r & \\ & & \bullet & \bullet & \bullet \\ & & & k_r & -(3k_r+k_{r,s}) \end{pmatrix} \quad (3\text{-}16)$$

As mentioned at the beginning of this section, many levels of decomposition and merging operations take place but most of those operations occur within each material (For detailed discussions see [7].) So as seen in the above equations and those to follow, the superscripts attached to vectors are at the last levels such as ($e$) and ($e$+1). Note that the superscript (0) indicates that the values given at the beginning of the S-PSM process do not change at the end of the process. Note also that the last two lines of Eq. (3-12) describe the correspondence between the variable names of each of the two materials (For the reasoning to derive such equations, see [7].).

We now get the following equations for the first and second material.

$$\left.\begin{array}{l} \mathbf{q}_1^{(e_2)} = \mathbf{f}_{q,1}^{(e_2+1)} - A_1^{(e_2+1)}\mathbf{q}_0^{(e_2)} - A_2^{(e_2+1)}\mathbf{q}_5^{(e_2)} \\ \mathbf{q}_4^{(e_2)} = \mathbf{f}_{q,2}^{(e_2+1)} - A_2^{(e_2+1)}\mathbf{q}_0^{(e_2)} - A_1^{(e_2+1)}\mathbf{q}_5^{(e_2)} \end{array}\right\} \quad (3\text{-}17)$$

$$\left.\begin{array}{l} \mathbf{r}_1^{(e_3)} = \mathbf{f}_{r,1}^{(e_3+1)} - A_1^{(e_3+1)}\mathbf{r}_0^{(e_3)} - A_2^{(e_3+1)}\mathbf{r}_5^{(e_3)} \\ \mathbf{r}_4^{(e_3)} = \mathbf{f}_{r,2}^{(e_3+1)} - A_2^{(f_3+1)}\mathbf{r}_0^{(e_3)} - A_1^{(e_3+1)}\mathbf{r}_5^{(e_3)} \end{array}\right\} \quad (3\text{-}18)$$

We apply the above procedure of derivation of Eqs. (3-12), (3-17) and (3-18) to the remaining boundaries. Then we get the following matrix-vector form of Eq. (3-19) for the three boundaries:

$$\begin{pmatrix} A_{p1}^{(0)} & I & O & O & & & & & & & & & & & & \\ A_1^{(e_1+1)} & I & O & A_2^{(e_1+1)} & & & & & & & & & & & & \\ A_2^{(e_1+1)} & O & I & A_1^{(e_1+1)} & & & & & & & & & & & & \\ O & O & k_pI & A_{p2}^{(0)} & k_{p,q}I & & & & & & & & & & & \\ & & & k_{p,q}I & A_{q2}^{(0)} & k_qI & O & O & & & & & & & & \\ & & & & A_1^{(e_2+1)} & I & O & A_2^{(e_2+1)} & & & & & & & & \\ & & & & A_2^{(e_2+1)} & O & I & A_1^{(e_2+1)} & & & & & & & & \\ & & & & O & O & k_qI & A_{q1}^{(0)} & k_{q,r}I & & & & & & & \\ & & & & & & & k_{q,r}I & A_{r1}^{(0)} & k_rI & O & O & & & & \\ & & & & & & & & A_1^{(e_3+1)} & I & O & A_2^{(e_3+1)} & & & & \\ & & & & & & & & A_2^{(e_3+1)} & O & I & A_1^{(e_3+1)} & & & & \\ & & & & & & & & O & O & k_rI & A_{r2}^{(0)} & k_{r,s}I & & & \\ & & & & & & & & & & & k_{r,s}I & A_{s2}^{(0)} & I & O & O \\ & & & & & & & & & & & & A_1^{(e_4+1)} & I & O & A_2^{(e_4+1)} \\ & & & & & & & & & & & & A_2^{(e_4+1)} & O & I & A_1^{(e_4+1)} \\ & & & & & & & & & & & & O & O & k_sI & A_{s1}^{(0)} \end{pmatrix} \begin{pmatrix} \mathbf{p}_0^{(e_1)} \\ \mathbf{p}_1^{(e_1)} \\ \mathbf{p}_4^{(e_1)} \\ \mathbf{p}_5^{(e_1)} \\ \mathbf{q}_5^{(e_2)} \\ \mathbf{q}_4^{(e_2)} \\ \mathbf{q}_1^{(e_2)} \\ \mathbf{q}_0^{(e_2)} \\ \mathbf{r}_0^{(e_3)} \\ \mathbf{r}_1^{(e_3)} \\ \mathbf{r}_4^{(e_3)} \\ \mathbf{r}_5^{(e_3)} \\ \mathbf{s}_5^{(e_4)} \\ \mathbf{s}_4^{(e_4)} \\ \mathbf{s}_1^{(e_4)} \\ \mathbf{s}_0^{(e_4)} \end{pmatrix} = \begin{pmatrix} \mathbf{f}_{p,0}^{(0)} \\ \mathbf{f}_{p,1}^{(e_1+1)} \\ \mathbf{f}_{p,2}^{(e_1+1)} \\ k_p\mathbf{f}_{p,n_p+1}^{(0)} \\ k_q\mathbf{f}_{q,n_q+1}^{(0)} \\ \mathbf{f}_{q,2}^{(e_2+1)} \\ \mathbf{f}_{q,1}^{(e_2+1)} \\ k_q\mathbf{f}_{q,0}^{(0)} \\ k_r\mathbf{f}_{r,0}^{(0)} \\ \mathbf{f}_{r,1}^{(e_3+1)} \\ \mathbf{f}_{r,2}^{(e_3+1)} \\ k_r\mathbf{f}_{r,n_r+1}^{(0)} \\ \mathbf{f}_{s,n_s+1}^{(0)} \\ \mathbf{f}_{s,2}^{(e_4+1)} \\ \mathbf{f}_{s,1}^{(e_4+1)} \\ k_s\mathbf{f}_{s,0}^{(0)} \end{pmatrix} - \begin{pmatrix} \mathbf{v}_0^{(0)} \\ 0 \\ 0 \\ 0 \\ 0 \\ 0 \\ 0 \\ 0 \\ 0 \\ 0 \\ 0 \\ 0 \\ \mathbf{w}_0^{(0)} \\ 0 \\ 0 \\ 0 \end{pmatrix}$$

(3-19)

### 3.3. System Decomposition and Partial Solutions for Each Subsystem/Material

We decompose Eq. (3-19) into four subsystems of equations that correspond to the four materials.

$$\begin{pmatrix} A_{p1}^{(0)} & I & O & O \\ A_1^{(e_1+1)} & I & O & A_2^{(e_1+1)} \\ A_2^{(e_1+1)} & O & I & A_1^{(e_1+1)} \\ O & O & k_pI & A_{p2}^{(0)} \end{pmatrix} \begin{pmatrix} \mathbf{p}_0^{(e_1)} \\ \mathbf{p}_1^{(e_1)} \\ \mathbf{p}_4^{(e_1)} \\ \mathbf{p}_5^{(e_1)} \end{pmatrix} = -\begin{pmatrix} \mathbf{v}_0^{(0)} \\ 0 \\ 0 \\ k_{p,q}\mathbf{q}_0^{(e_2)} \end{pmatrix} + \begin{pmatrix} \mathbf{f}_{p,0}^{(0)} \\ \mathbf{f}_{p,1}^{(e_1+1)} \\ \mathbf{f}_{p,2}^{(e_1+1)} \\ k_p\mathbf{f}_{p,n_p+1}^{(0)} \end{pmatrix} \quad (3\text{-}20)$$

$$\begin{pmatrix} A_{q2}^{(0)} & k_qI & O & O \\ A_1^{(e_2+1)} & I & O & A_2^{(e_2+1)} \\ A_2^{(e_2+1)} & O & I & A_1^{(e_2+1)} \\ O & O & k_qI & A_{q1}^{(0)} \end{pmatrix} \begin{pmatrix} \mathbf{q}_5^{(e_2)} \\ \mathbf{q}_4^{(e_2)} \\ \mathbf{q}_1^{(e_2)} \\ \mathbf{q}_0^{(e_2)} \end{pmatrix} = -\begin{pmatrix} k_{q,r}\mathbf{r}_0^{(e_3)} \\ 0 \\ 0 \\ k_{p,q}\mathbf{p}_5^{(e_1)} \end{pmatrix} + \begin{pmatrix} k_q\mathbf{f}_{q,n_q+1}^{(0)} \\ \mathbf{f}_{q,2}^{(e_2+1)} \\ \mathbf{f}_{q,1}^{(e_2+1)} \\ k_q\mathbf{f}_{q,0}^{(0)} \end{pmatrix} \quad (3\text{-}21)$$

$$\begin{pmatrix} A_{r1}^{(0)} & k_rI & O & O \\ A_1^{(e_3+1)} & I & O & A_2^{(e_3+1)} \\ A_2^{(e_3+1)} & O & I & A_1^{(e_3+1)} \\ O & O & k_rI & A_{r2}^{(0)} \end{pmatrix} \begin{pmatrix} \mathbf{r}_0^{(e_3)} \\ \mathbf{r}_1^{(e_3)} \\ \mathbf{r}_4^{(e_3)} \\ \mathbf{r}_5^{(e_3)} \end{pmatrix} = -\begin{pmatrix} k_{q,r}\mathbf{q}_5^{(e_2)} \\ 0 \\ 0 \\ k_{r,s}\mathbf{s}_0^{(e_4)} \end{pmatrix} + \begin{pmatrix} k_r\mathbf{f}_{r,0}^{(0)} \\ \mathbf{f}_{r,1}^{(e_3+1)} \\ \mathbf{f}_{r,2}^{(e_3+1)} \\ k_r\mathbf{f}_{r,n_r+1}^{(0)} \end{pmatrix} \quad (3\text{-}22)$$

$$\begin{pmatrix} A_{s2}^{(0)} & I & O & O \\ A_1^{(e_4+1)} & I & O & A_2^{(e_4+1)} \\ A_2^{(e_4+1)} & O & I & A_1^{(e_4+1)} \\ O & O & k_sI & A_{s1}^{(0)} \end{pmatrix} \begin{pmatrix} \mathbf{s}_5^{(e_4)} \\ \mathbf{s}_4^{(e_4)} \\ \mathbf{s}_1^{(e_4)} \\ \mathbf{s}_0^{(e_4)} \end{pmatrix} = -\begin{pmatrix} \mathbf{w}_0^{(0)} \\ 0 \\ 0 \\ k_{r,s}\mathbf{r}_5^{(e_3)} \end{pmatrix} + \begin{pmatrix} \mathbf{f}_{s,n_s+1}^{(0)} \\ \mathbf{f}_{s,2}^{(e_4+1)} \\ \mathbf{f}_{s,1}^{(e_4+1)} \\ k_s\mathbf{f}_{s,0}^{(0)} \end{pmatrix} \quad (3\text{-}23)$$

We denote the inverse matrix of each coefficient matrix of the above equations by the matrix of 16 submatrices of the form $B^u$ as shown below. The partial solutions for each subsystem are given as follows. (For the correspondence between the blocks of the inverse matrix above and the B matrix below, see [7].)

$$\begin{pmatrix} \mathbf{p}_0^{(e_1)} \\ \mathbf{p}_1^{(e_1)} \\ \mathbf{p}_4^{(e_1)} \\ \mathbf{p}_5^{(e_1)} \end{pmatrix} = \begin{pmatrix} B_{1,1}^p & B_{1,2}^p & B_{1,3}^p & B_{1,4}^p \\ B_{2,1}^p & B_{2,2}^p & B_{2,3}^p & B_{2,4}^p \\ B_{3,1}^p & B_{3,2}^p & B_{3,3}^p & B_{3,4}^p \\ B_{4,1}^p & B_{4,2}^p & B_{4,3}^p & B_{4,4}^p \end{pmatrix} \begin{pmatrix} \mathbf{v}_0^{(0)} \\ 0 \\ 0 \\ k_{p,q}\mathbf{q}_0^{(e_2)} \end{pmatrix} + \begin{pmatrix} \mathbf{f}_{p,0}^{(0)} \\ \mathbf{f}_{p,1}^{(e_1+1)} \\ \mathbf{f}_{p,2}^{(e_1+1)} \\ k_p\mathbf{f}_{p,n_p+1}^{(0)} \end{pmatrix} \quad (3\text{-}24)$$

$$\begin{pmatrix} \mathbf{q}_5^{(e_2)} \\ \mathbf{q}_4^{(e_2)} \\ \mathbf{q}_1^{(e_2)} \\ \mathbf{q}_0^{(e_2)} \end{pmatrix} = \begin{pmatrix} B_{1,1}^q & B_{1,2}^q & B_{1,3}^q & B_{1,4}^q \\ B_{2,1}^q & B_{2,2}^q & B_{2,3}^q & B_{2,4}^q \\ B_{3,1}^q & B_{3,2}^q & B_{3,3}^q & B_{3,4}^q \\ B_{4,1}^q & B_{4,2}^q & B_{4,3}^q & B_{4,4}^q \end{pmatrix} \begin{pmatrix} k_{q,r}\mathbf{r}_0^{(e_3)} \\ 0 \\ 0 \\ k_{p,q}\mathbf{p}_5^{(e_1)} \end{pmatrix} + \begin{pmatrix} k_q\mathbf{f}_{q,n_q+1}^{(0)} \\ \mathbf{f}_{q,2}^{(e_2+1)} \\ \mathbf{f}_{q,1}^{(e_2+1)} \\ k_q\mathbf{f}_{q,0}^{(0)} \end{pmatrix} \quad (3\text{-}25)$$

$$\begin{pmatrix} \mathbf{r}_0^{(e_3)} \\ \mathbf{r}_1^{(e_3)} \\ \mathbf{r}_4^{(e_3)} \\ \mathbf{r}_5^{(e_3)} \end{pmatrix} = \begin{pmatrix} B_{1,1}^r & B_{1,2}^r & B_{1,3}^r & B_{1,4}^r \\ B_{2,1}^r & B_{2,2}^r & B_{2,3}^r & B_{2,4}^r \\ B_{3,1}^r & B_{3,2}^r & B_{3,3}^r & B_{3,4}^r \\ B_{4,1}^r & B_{4,2}^r & B_{4,3}^r & B_{4,4}^r \end{pmatrix} \begin{pmatrix} k_{q,r}\mathbf{q}_5^{(e_2)} \\ 0 \\ 0 \\ k_{r,s}\mathbf{s}_0^{(e_4)} \end{pmatrix} + \begin{pmatrix} k_r\mathbf{f}_{r,0}^{(0)} \\ \mathbf{f}_{r,1}^{(e_3+1)} \\ \mathbf{f}_{r,2}^{(e_3+1)} \\ k_r\mathbf{f}_{r,n_r+1}^{(0)} \end{pmatrix} \quad (3\text{-}26)$$

$$\begin{pmatrix} \mathbf{s}_5^{(e_4)} \\ \mathbf{s}_4^{(e_4)} \\ \mathbf{s}_1^{(e_4)} \\ \mathbf{s}_0^{(e_4)} \end{pmatrix} = \begin{pmatrix} B_{1,1}^s & B_{1,2}^s & B_{1,3}^s & B_{1,4}^s \\ B_{2,1}^s & B_{2,2}^s & B_{2,3}^s & B_{2,4}^s \\ B_{3,1}^s & B_{3,2}^s & B_{3,3}^s & B_{3,4}^s \\ B_{4,1}^s & B_{4,2}^s & B_{4,3}^s & B_{4,4}^s \end{pmatrix} \begin{pmatrix} \mathbf{w}_0^{(0)} \\ 0 \\ 0 \\ k_{r,s}\mathbf{r}_5^{(e_3)} \end{pmatrix} + \begin{pmatrix} \mathbf{f}_{s,n_s+1}^{(0)} \\ \mathbf{f}_{s,2}^{(e_4+1)} \\ \mathbf{f}_{s,1}^{(e_4+1)} \\ k_s\mathbf{f}_{s,0}^{(0)} \end{pmatrix} \quad (3\text{-}27)$$

### 3.4. Merging of Partial Solutions for Each Pair of Adjacent Materials

For the pair of adjacent materials $p$ and $q$, we extract two equations for variables $\mathbf{p}_5^{(e_1)}$ and $\mathbf{q}_0^{(e_2)}$ from Eqs. (3-24) and (3-25), respectively, and merge them to get Eq. (3-28). Similarly, for the $r$, $s$ pair, we merge the two equations with respect to $\mathbf{r}_5^{(e_3)}$ and $\mathbf{s}_0^{(e_4)}$ from Eqs. (3-26) and (3-27), and derive Eq. (3-29).

$$\begin{pmatrix} I & O & k_{p,q}B_{1,4}^p & O \\ O & I & k_{p,q}B_{4,4}^p & O \\ O & k_{p,q}B_{4,4}^q & I & O \\ O & k_{p,q}B_{1,4}^q & O & I \end{pmatrix} \begin{pmatrix} \mathbf{p}_0^{(e_1)} \\ \mathbf{p}_5^{(e_1)} \\ \mathbf{q}_0^{(e_2)} \\ \mathbf{q}_5^{(e_2)} \end{pmatrix} = -\begin{pmatrix} B_{1,1}^p\mathbf{v}_0^{(0)} \\ B_{4,1}^p\mathbf{v}_0^{(0)} \\ k_{q,r}B_{4,1}^q\mathbf{r}_0^{(e_3)} \\ k_{q,r}B_{1,1}^q\mathbf{r}_0^{(e_3)} \end{pmatrix} + \begin{pmatrix} \mathbf{ff}_{p0} \\ \mathbf{ff}_{p5} \\ \mathbf{ff}_{q0} \\ \mathbf{ff}_{q5} \end{pmatrix} \quad (3\text{-}28)$$

$$\begin{pmatrix} I & O & k_{r,s}B_{1,4}^r & O \\ O & I & k_{r,s}B_{4,4}^r & O \\ O & k_{r,s}B_{4,4}^s & I & O \\ O & k_{r,s}B_{1,4}^s & O & I \end{pmatrix} \begin{pmatrix} \mathbf{r}_0^{(e_3)} \\ \mathbf{r}_5^{(e_3)} \\ \mathbf{s}_0^{(e_4)} \\ \mathbf{s}_5^{(e_4)} \end{pmatrix} = -\begin{pmatrix} k_{q,r}B_{1,1}^r\mathbf{q}_5^{(e_2)} \\ k_{q,r}B_{4,1}^r\mathbf{q}_5^{(e_2)} \\ B_{4,1}^s\mathbf{w}_0^{(0)} \\ B_{1,1}^s\mathbf{w}_0^{(0)} \end{pmatrix} + \begin{pmatrix} \mathbf{ff}_{r0} \\ \mathbf{ff}_{r5} \\ \mathbf{ff}_{s0} \\ \mathbf{ff}_{s5} \end{pmatrix} \quad (3\text{-}29)$$

where

$$\begin{aligned}
\mathbf{ff}_{p0} &= B_{1,1}^p\mathbf{f}_{p,0}^{(0)} + B_{1,2}^p\mathbf{f}_{p,1}^{(e_1+1)} + B_{1,3}^p\mathbf{f}_{p,2}^{(e_1+1)} + k_pB_{1,4}^p\mathbf{f}_{p,n_p+1}^{(0)} \\
\mathbf{ff}_{p5} &= B_{4,1}^p\mathbf{f}_{p,0}^{(0)} + B_{4,2}^p\mathbf{f}_{p,1}^{(e_1+1)} + B_{4,3}^p\mathbf{f}_{p,2}^{(e_1+1)} + k_pB_{4,4}^p\mathbf{f}_{p,n_p+1}^{(0)} \\
\mathbf{ff}_{q0} &= k_qB_{4,1}^q\mathbf{f}_{q,n_q+1}^{(0)} + B_{4,2}^q\mathbf{f}_{q,2}^{(e_2+1)} + B_{4,3}^q\mathbf{f}_{q,1}^{(e_2+1)} + k_qB_{4,4}^q\mathbf{f}_{q,0}^{(0)} \\
\mathbf{ff}_{q5} &= k_qB_{1,1}^q\mathbf{f}_{q,n_q+1}^{(0)} + B_{1,2}^q\mathbf{f}_{q,2}^{(e_2+1)} + B_{1,3}^q\mathbf{f}_{q,1}^{(e_2+1)} + k_qB_{1,4}^q\mathbf{f}_{q,0}^{(0)} \\
\mathbf{ff}_{r0} &= k_rB_{1,1}^r\mathbf{f}_{r,0}^{(0)} + B_{1,2}^r\mathbf{f}_{r,1}^{(e_3+1)} + B_{1,3}^r\mathbf{f}_{r,2}^{(e_3+1)} + k_rB_{1,4}^r\mathbf{f}_{r,n_r+1}^{(0)} \\
\mathbf{ff}_{r5} &= k_rB_{4,1}^r\mathbf{f}_{r,0}^{(0)} + B_{4,2}^r\mathbf{f}_{r,1}^{(e_3+1)} + B_{4,3}^r\mathbf{f}_{r,2}^{(e_3+1)} + k_rB_{4,4}^r\mathbf{f}_{r,n_r+1}^{(0)} \\
\mathbf{ff}_{s0} &= B_{4,1}^s\mathbf{f}_{s,n_s+1}^{(0)} + B_{4,2}^s\mathbf{f}_{s,2}^{(e_4+1)} + B_{4,3}^s\mathbf{f}_{s,1}^{(e_4+1)} + k_sB_{4,4}^s\mathbf{f}_{s,0}^{(0)} \\
\mathbf{ff}_{s5} &= B_{1,1}^s\mathbf{f}_{s,n_s+1}^{(0)} + B_{1,2}^s\mathbf{f}_{s,2}^{(e_4+1)} + B_{1,3}^s\mathbf{f}_{s,1}^{(e_4+1)} + k_sB_{1,4}^s\mathbf{f}_{s,0}^{(0)}
\end{aligned} \quad (3\text{-}30)$$

As both of the above coefficient matrices are of special structure, their inverse matrices are expressed as shown below with two full middle columns of submatrices of the forms $B^{pq}$ and $B^{rs}$, respectively. The partial solutions for each pair of materials are now obtained as follows. (See [7].)

$$\begin{pmatrix} \mathbf{p}_0^{(e_1)} \\ \mathbf{p}_5^{(e_1)} \\ \mathbf{q}_0^{(e_2)} \\ \mathbf{q}_5^{(e_2)} \end{pmatrix} = \begin{pmatrix} I & B_{1,1}^{pq} & B_{2,1}^{pq} & O \\ O & B_{1,2}^{pq} & B_{2,2}^{pq} & O \\ O & B_{1,3}^{pq} & B_{2,3}^{pq} & O \\ O & B_{1,4}^{pq} & B_{2,4}^{pq} & I \end{pmatrix} \begin{pmatrix} B_{1,1}^p\mathbf{v}_0^{(0)} \\ B_{4,1}^p\mathbf{v}_0^{(0)} \\ k_{q,r}B_{4,1}^q\mathbf{r}_0^{(e_3)} \\ k_{q,r}B_{1,1}^q\mathbf{r}_0^{(e_3)} \end{pmatrix} + \begin{pmatrix} \mathbf{ff}_{p0} \\ \mathbf{ff}_{p5} \\ \mathbf{ff}_{q0} \\ \mathbf{ff}_{q5} \end{pmatrix} \quad (3\text{-}31)$$

$$\begin{pmatrix} \mathbf{r}_0^{(e_3)} \\ \mathbf{r}_5^{(e_3)} \\ \mathbf{s}_0^{(e_4)} \\ \mathbf{s}_5^{(e_4)} \end{pmatrix} = \begin{pmatrix} I & B_{1,1}^{rs} & B_{2,1}^{rs} & O \\ O & B_{1,2}^{rs} & B_{2,2}^{rs} & O \\ O & B_{1,3}^{rs} & B_{2,3}^{rs} & O \\ O & B_{1,4}^{rs} & B_{2,4}^{rs} & I \end{pmatrix} \begin{pmatrix} k_{q,r}B_{1,1}^r\mathbf{q}_5^{(e_2)} \\ k_{q,r}B_{4,1}^r\mathbf{q}_5^{(e_2)} \\ B_{4,1}^s\mathbf{w}_0^{(0)} \\ B_{1,1}^s\mathbf{w}_0^{(0)} \end{pmatrix} + \begin{pmatrix} \mathbf{ff}_{r0} \\ \mathbf{ff}_{r5} \\ \mathbf{ff}_{s0} \\ \mathbf{ff}_{s5} \end{pmatrix} \quad (3\text{-}32)$$

### 3.5. Final Solutions

Finally, for the two pairs of materials $p$ and $q$, and of $r$ and s, we extract equations with respect to variables $\mathbf{q}_5^{(e_2)}$ and $\mathbf{r}_0^{(e_3)}$ from Eqs. (3-31) and (3-32). Their merging then produces the following equation.

$$\begin{pmatrix} I & k_{q,r}\left(B_{2,4}^{pq}B_{4,1}^q + B_{1,1}^q\right) \\ k_{q,r}\left(B_{1,1}^{rs}B_{4,1}^r + B_{1,1}^r\right) & I \end{pmatrix} \begin{pmatrix} \mathbf{q}_5^{(e_2)} \\ \mathbf{r}_0^{(e_3)} \end{pmatrix} = -\begin{pmatrix} B_{1,4}^{pq}B_{4,1}^p\mathbf{v}_0^{(0)} \\ B_{2,1}^{rs}B_{4,1}^s\mathbf{w}_0^{(0)} \end{pmatrix} + \begin{pmatrix} \mathbf{ff}_{pq} \\ \mathbf{ff}_{rs} \end{pmatrix} \quad (3\text{-}33)$$

where

$$\left.\begin{array}{l}\mathbf{ff}_{pq} = B_{1,4}^{pq}\mathbf{ff}_{p5} + B_{2,4}^{pq}\mathbf{ff}_{q0} + \mathbf{ff}_{q5} \\ \mathbf{ff}_{rs} = \mathbf{ff}_{r0} + B_{1,1}^{rs}\mathbf{ff}_{r5} + B_{2,1}^{rs}\mathbf{ff}_{s0}\end{array}\right\} \quad (3\text{-}34)$$

We then take the inverse matrix of the coefficient matrix of Eq. (3-33) and derive the final solutions for variables $\mathbf{q}_5^{(e_2)}$ and $\mathbf{r}_0^{(e_3)}$ as follows. (See [7].)

$$\left.\begin{array}{l}\mathbf{q}_5^{(e_2)} = -D_1(B_{1,4}^{pq}(B_{4,1}^p\mathbf{v}_0^{(0)})) - D_2(B_{2,1}^{rs}(B_{4,1}^s\mathbf{w}_0^{(0)})) + D_1\mathbf{ff}_{pq} + D_2\mathbf{ff}_{rs} \\ \mathbf{r}_0^{(e_3)} = -D_3(B_{1,4}^{pq}(B_{4,1}^p\mathbf{v}_0^{(0)})) - D_4(B_{2,1}^{rs}(B_{4,1}^s\mathbf{w}_0^{(0)})) + D_3\mathbf{ff}_{pq} + D_4\mathbf{ff}_{rs}\end{array}\right\}$$
(3-35)

where

$$\left.\begin{array}{l}D_1 = [I - k_{q,r}k_{r,r}(B_{2,4}^{pq}B_{4,1}^q + B_{1,1}^q)(B_{1,1}^{rs}B_{4,1}^r + B_{1,1}^r)]^{-1} \\ D_2 = -k_{q,r}D_1(B_{2,4}^{pq}B_{4,1}^q + B_{1,1}^q) \\ D_4 = [I - k_{q,r}k_{r,r}(B_{1,1}^{rs}B_{4,1}^r + B_{1,1}^r)(B_{2,4}^{pq}B_{4,1}^q + B_{1,1}^q)]^{-1} \\ D_3 = -k_{q,r}D_4(B_{1,1}^{rs}B_{4,1}^r + B_{1,1}^r)\end{array}\right\}$$
(3-36)

### 3.6. Back Substitution

The above solutions are now substituted into Eqs. (3-31) and (3-32) to find solutions for $\mathbf{p}_0^{(e_1)}$, $\mathbf{p}_5^{(e_1)}$, $\mathbf{q}_0^{(e_2)}$, $\mathbf{r}_5^{(e_3)}$, $\mathbf{s}_0^{(e_4)}$ and $\mathbf{s}_5^{(e_4)}$. These solutions are then back substituted into Eqs. (3-24), (3-25), (3-26) and (3-27) and the solutions for the remaining variables are obtained. It should be noted that we need one more step to find solutions for the variables associated with each material. Along a similar line of equation derivations given above, this can be done by way of repeated substitutions of the values for relevant variables into certain equations. (See [7] for more detail.)

### 4. Experiments

We applied the technique described above to two dimensional steady-state heat conduction analysis at the chip level of four layer materials mounted on a mother board. They are a VLSI chip, a thermal interface material (*e.g.*, adhesive), a heat spreader, and a heat sink as shown in Fig. 2.1 The parameter values, such as the thermal conductivity and the depth and width of each material given in the figure are those used in the industrial design settings.

In our example, we assume that the active silicon bulk is the only heat source and that no Joule heat is generated from the metal layer of wire. So the temperature of the heat source, which is the value of junction temperature of $T_j = 120°C$, and the ambient temperature of $T_{amb} = 55°C$ are set as the boundary conditions at the bottom and top edges, respectively, of the region analyzed. These values are also used in the industrial design settings. As for the left and right boundary conditions, we set $T_{right} = T_{left} = 20°C$ as the reference (room) temperatures for our analysis. This room temperature of 20°C was also used in reference [14]. Reference [10] used the reference temperature of 0°C

To evaluate the performance of our technique, we developed a C program for solving the four Laplace equations described above. Although it took more time to develop this C code than expected, the resultant program was rather simple. We ran it on an IBM Z Pro (OS Linux, CPU Xeon 3.60GHz, memory 32GB). We also ran a commercial Laplace equation solver software, called Raphael [11]. Figure 4.1 depicts the temperature distributions obtained by our program and Raphael. Both figures resemble very close to each other. This implies that our technique produces correct results.

Table I shows the CPU times required and the residuals produced by our program and Raphael. The results demonstrate that for the largest grid, our program ran 5.8 and 8.9 times faster while keeping smaller residuals by 5 and 1 order of magnitudes, than Raphael.

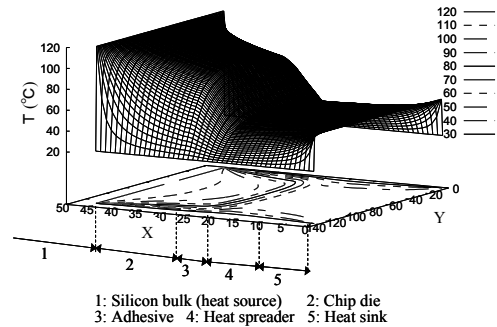

Case 1: Results by our program

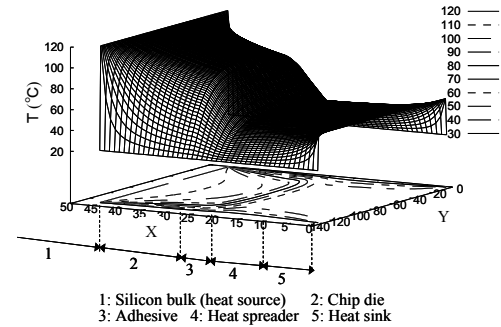

Case 2: Results by Raphael

**Fig. 4.1: Temperature distributions of steady-state heat conduction for four layers of materials obtained by our program and Raphael.**

**Table I: Processing CPU times of and residuals by our program and Raphael for difference arrangements of grid points.**

| Number of grid points | | | | | | S-PSM | | Raphael (EPS=1.0E-6) | | Raphael (EPS=1.0E-10) | |
|---|---|---|---|---|---|---|---|---|---|---|---|
| h | g | $n_p$ | $n_q$ | $n_r$ | $n_s$ | CPU(sec) | Residual | CPU(sec) | Residual | CPU(sec) | Residual |
| 128 | 44 | 8 | 8 | 4 | 16 | 0.05 | 1.3E-13 | 0.26 | 9.5E-07 | 0.32 | 9.3E-11 |
| 256 | 80 | 16 | 16 | 8 | 32 | 0.18 | 5.1E-13 | 1.98 | 9.9E-07 | 2.78 | 9.4E-11 |
| 512 | 152 | 32 | 32 | 16 | 64 | 1.7 | 2.1E-12 | 13.2 | 9.9E-07 | 19.5 | 9.9E-11 |
| 1,024 | 296 | 64 | 64 | 32 | 128 | 15.4 | 8.9E-12 | 89.9 | 9.7E-07 | 136.6 | 9.9E-11 |

## 5. Conclusions

We have proposed a new technique of thermal analysis for multi-layer VLSI chips. It starts with the modeling of multiple layers of materials of different heat conductivities by a set of Laplace equations. It then discretizes the equations by FDM and finally uses S-PSM to obtain numerical solutions for the resulting systems of linear equations. We applied our technique to steady-state heat conduction analysis at the chip level for a realistic case of four layer materials. The experimental results showed 5.8 and 8.9 times speed-up and simultaneous residual improvement by 5 and 1 order of magnitude, respectively, by our program over the commercial tool, called Raphael.

In our experiments, we assumed that the temperature of the heat source was given. The generation of the heat may be expressed by a Poisson equation and its solution could easily be computed by S-PSM. Extension of our work to the process of a Poisson equation and other cases, such as three dimensional and transient heat conduction analysis are under way. We can also easily extend our work to the case of complex shapes and boundary conditions of materials.

## 6. Acknowledgment

The authors would like to express their gratitude to Dr. T. Hiramoto for his introduction of S-PSM to them. They are also grateful to anonymous reviewers for their suggestions